\documentclass[letterpaper,twocolumn,american,showpacs,pre,aps,
superscriptaddress ]{revtex4-2}
\usepackage[latin1]{inputenc}
\usepackage{bm}
\usepackage{multirow}
\usepackage{amssymb}
\usepackage{amsbsy}
\usepackage{amsmath}
\usepackage{stmaryrd}
\usepackage{graphicx}
\usepackage{subfigure}
\usepackage{epsfig}
\usepackage{hyperref}
\makeatletter
\usepackage{pifont}
\makeatother

\begin{document}

\title{Estimation of equilibration time scales from nested fraction approximations}

\author{Christian Bartsch}
\email{cbartsch@uos.de}
\affiliation{Department of Mathematics/Computer Science/Physics, University of
Osnabr\"uck, D-49076 Osnabr\"uck, Germany}

\author{Anatoly Dymarsky}
\affiliation{Department of Physics, University of Kentucky, Lexington, Kentucky, USA, 40506}

\author{Mats H. Lamann}
\affiliation{Department of Mathematics/Computer Science/Physics, University of
Osnabr\"uck, D-49076 Osnabr\"uck, Germany}

\author{Jiaozi Wang}
\affiliation{Department of Mathematics/Computer Science/Physics, University of
Osnabr\"uck, D-49076 Osnabr\"uck, Germany}

\author{Robin Steinigeweg}
\affiliation{Department of Mathematics/Computer Science/Physics, University of
Osnabr\"uck, D-49076 Osnabr\"uck, Germany}

\author{Jochen Gemmer}
\affiliation{Department of Mathematics/Computer Science/Physics, University of
Osnabr\"uck, D-49076 Osnabr\"uck, Germany}

\date{\today}

\begin{abstract}
We consider an autocorrelation function of a quantum mechanical system through the lens of the so-called recursive method, by iteratively evaluating Lanczos coefficients, 
or solving a system of coupled differential equations in the Mori formalism. We first show that both methods are mathematically equivalent, each offering certain practical advantages. We then propose an approximation scheme to evaluate the autocorrelation function, and use it 
to estimate the equilibration time $\tau$ for the observable in question. With only a handful of Lanczos coefficients as the input, this scheme yields an accurate order of magnitude estimate of $\tau$, matching state-of-the-art numerical approaches. We develop a simple numerical scheme to estimate the precision of our method. We test our approach using several numerical examples exhibiting different relaxation dynamics.
Our findings provide a new practical way to quantify the
equilibration time of isolated quantum systems, a question which is both crucial and notoriously difficult.
\end{abstract}

\maketitle

\section{Introduction}

How do many-body quantum systems approach equilibrium? The question to quantify this behavior goes back to the advent of quantum
mechanics. Considerable progress has been made in the past decades, with concepts like typicality
and the eigenstate thermalization hypothesis having been discovered (and
sometimes rediscovered) \cite{gogolin2016}. However, even the question if an
expectation value $\langle O(t) \rangle = \text{Tr}\{ O(t)\rho \}$ will reach a
certain equilibration value after some time $T$ and never depart from it thereafter (until
the Poincar\'e recurrence time, which is usually parametrically much larger than $T$), for any concrete 
few-body observable $O$, Hamiltonian $H$, and an initial state $\rho$, can not currently be 
answered with certainty. An often used concept in this context is
the ``equilibration on average'' \cite{reimann2008}. Here, schematically, the
frequency of instants in time at which $\langle O(t)\rangle$ significantly
deviates from its temporal average, is considered. There are well developed rigorous theorems, asserting this frequency is
very small for most practical situations. However, 
it is very hard (c.f.\ below) to put a bound on the time interval for which this
frequency statement applies. Available rigorous bounds are usually too conservative, exceeding physically observed thermalization time by many orders of magnitude. 
Furthermore, very odd dynamics that are nevertheless in full
accord with the principle of equilibration on average have been demonstrated
\cite{knipschild2020}. While it has been shown that for random
Hamiltonians or observables, these equilibration times are typically very short
\cite{goldstein2013,malabarba2014,reimann2016,vinayak2012,brandao2012,
masanes2013,goldstein2015,alhambra2020}, it is easy to construct setups for
which they are exceedingly long \cite{goldstein2013,malabarba2014}. There is an extensive literature establishing bounds on equilibration times, both upper bounds
\cite{garcia2017,alhambra2020}, as well as lower bounds (often called
speed limits) \cite{hamazaki2022}. While these attempts certainly advance the
field, many problems limiting their practical applicability remain
\cite{heveling2020}.

In this paper we take a different approach to estimate equilibration time of an isolated quantum system. We focus
on the dynamics of autocorrelation functions at infinite temperature, i.e.,
$C(t)=\text{Tr}\{ O(t)O \}$. The latter is in close relation with the dynamics
of the expectation values $\langle O(t) \rangle$ for a great variety of 
initial out of equilibrium states $\rho$ \cite{richter2019,richter2019-1}. The time dynamics of $C(t)$ can be expressed using the recursion method or in the Mori formalism. Both approaches parametrize $C(t)$ in terms of a sequence of positive numbers $b_n$, called the Lanczos
coefficients, which are defined by the pair $O$, $H$. We show that these
numbers are the same in both pictures and further elaborate on the relation
between these two methods. We then propose an approximation scheme 
 and define $C^R(t)$, an approximation to $C(t)$, defined in terms of the first $R$ Lanczos coefficients. Technically, $C^R(t)$ is defined by leaving first $R$ coefficients $b_n$ intact and declaring $b_n=b_R$ for $n\geq R$. 
Strictly speaking $C^R(t)$ becomes $C(t)$ only in the limit $R\rightarrow \infty$, but importantly it gives a reasonable approximation for a wide range of times already for  modest values of $R$. We develop an indirect quantitative test of accuracy, which can signal that a reasonable approximation of $\tau$ is reached without the need to know the actual  $C(t)$. 
We then use $C^R(t)$ to estimate the thermalization time for a number of standard
observables and Hamiltonians and find that our method yields reasonable accuracy across the board already with $R < 10$.
Since computing first $10$ or so Lanczos coefficients for local 
observables and Hamiltonians with local interactions is usually a straightforward and simple numerical task, our approach readily provides a practical way to 
estimate the equilibration time for many quantum systems and observables. 
To summarize, our estimate is not a rigorous 
bound, but turns out to be very reasonable in all
considered examples.

This manuscript is organized as follows. First, in Sec.\ \ref{Lanczos}, we 
introduce the recursion method to evaluate the autocorrelation functions. Afterwards, 
in Sec.\ \ref{Mori}, we give a brief overview of the so-called Mori formalism 
for correlation functions and show the relation to the Lanczos formulation.
We then present a scheme to obtain an approximation for the 
autocorrelation function within the Mori picture, define suitable equilibration times
and establish a convergence criterion based on a pertinent area measure in Sec.\ \ref{Approx}.
Finally, we numerically test our approach for different types of quantum dynamics 
in Sec.\ \ref{Numerics}.

\section{Recursive method and\\  Lanczos coefficients}
\label{Lanczos}
Lanczos coefficients emerge as a part of the so-called recursion method. The object of interest is the 
autocorrelation functions of the form
\begin{equation}
C(t) = \text{Tr}\{ O(t)O \}
\end{equation}
with some pertinent observable $O$. Here, $O(t)$ denotes the time dependence
in the Heisenberg picture, $O(t) = e^{iHt}\, O\, e^{-iHt}$, induced by
corresponding Hamiltonian $H$ ($\hbar$ set to $1$). 

In Liouville space, i.e., the Hilbert space of observables, the elements $O$
can be denoted as states $| O)$. The time evolution is then induced
by the Liouville superoperator defined as $\mathcal{L} |O) = |[H,O])$.
A suitable inner product is given by $(O_1 | O_2)= \text{Tr}\{ O_1^{\dagger}
O_2\}$, which in turn defines a norm $\| O \| = \sqrt{(O|O)}$. The correlation
function $C(t)$ may then be written as $C(t) = (O|e^{i\mathcal{L}t}|O)$.

Now, one may iteratively construct a set of observables $O_n$ starting with the
``seed'' $O_0=O$, where $O$ is assumed to be normalized, i.e., $(O_0|O_0) = 1$.
We set $b_1 = \| \mathcal{L} O_0 \|$ and $|O_1) = \mathcal{L}|O_0)  / b_1$.
One can now employ the (infinite) iteration scheme 
\begin{eqnarray}
|Q_n) &=& \mathcal{L}|O_{n-1}) - b_{n-1} |O_{n-2}) \ , \\
b_n &=& \| Q_n\| \ , \\
|O_n) &=& |Q_n)  /  b_n \ ,
\end{eqnarray}
where $\{|O_n )\}$ constitute the Krylov basis and $b_n$ are the Lanczos
coefficients, which are real and positive and the crucial constituent in our
analysis \cite{cyrot1967}.

Using the Lanczos algorithm, one may cast the time evolution of
the components of a vector $\vec{x}$,  defined as $x_n (t) = i^{-n}
(O_n|O(t))$, into the form
\begin{equation}
 \dot{\vec{x}} = L\, \vec{x}
 \label{dynlancz}
\end{equation}
with the matrix $L$,
\begin{equation}
L =
\begin{pmatrix}
0 & -b_1 & 0 & 0 &\cdots \\
b_1 & 0 & -b_2 & 0 &\cdots\\
0 & b_2 & 0 & -b_3 & \cdots\\
\vdots  & \vdots  & \vdots & \vdots & \ddots
\end{pmatrix} \, ,
\end{equation}
which is similar to the standard Schr\"odinger equation. Then, the first
component of $\vec{x}$ is the correlation function of interest, $x_0 (t)= C(t)$.
The initial condition is always given by $x_n (0) = \delta_{0n}$.

Note that $L$ is here an anti-Hermitian tridiagonal matrix, but the
components $x_n$ are all real numbers which is very convenient for the following
analysis. 

For clarity, the above set of equations may also be written as
\begin{eqnarray}
\dot{x}_0 (t) &=& - b_1 x_1 (t) \, , \\
\dot{x}_n (t) &=& b_n x_{n-1} (t) - b_{n+1} x_{n+1} (t) \, , \quad n \geq 1 \, .
\label{dynlanczchain}
\end{eqnarray}
That is, the dynamics of the correlation function $C(t)$ may be calculated
as occupation amplitude of the ``first'' site in a one-dimensional
non-interacting hopping model without specifying any details of $H$ and $A$.

In the following, we assume that the final Lanczos coefficient $b_{m+1}$ is equal to
$0$ (for possibly large $m$). This is always the case for systems with a finite-dimensional Hilbert space. Doing so closes the infinite 
set of equations in the Lanczos formulation (\ref{dynlancz}), which in turn
allows us performing the Laplace transform of the now finite set of
equations to arrive at the following finite  ``continued'' fraction expansion.
Say, when $m=3$ and $b_4 = 0$ we have explicitly 
\begin{equation}
\label{LLT}
\tilde{x}_0 (s) = \frac{1}{s+\frac{b_1^2}{s+\frac{b_2^2}{s+\frac{b_3^2}{s}}}}
\end{equation}
Here, $\tilde{x}_0 (s)$ is the Laplace transform of $x_0 (t)$.

The continued fraction representations of the equations of motion Eq.(\ref{dynlanczchain}) was given in \cite{haydock1980}.

In practice the complexity of evaluating $b_n$ numerically quickly grows with $n$, such that only a handful first $b_n$ are usually available. 
For a local operator $O$ and a Hamiltonian with local interactions the first coefficients $b_n$, up to $n$ of order of the system size, are system size independent. This is an important observation allowing us later to estimate the thermalization time of $C(t)$ in the thermalization limit, when the system is taken to infinity. 

\section{Mori formulation of dynamics}
\label{Mori}

In the previous section we presented the description the
correlation function in terms of the recursive method and the Lanczos algorithm. 
There is an alternative approach of Mori 
(\cite{mori1965,joslin1986}), with the time evolution specified by a set of functions 
$C_n (t)$ satisfying a coupled series of Volterra equations,
\begin{equation}
\dot{C}_n (t) = - \Delta_{n+1}^2 \int_{0}^{t} C_{n+1}(t-t') C_{n} (t') dt' \ ,
\label{mori}
\end{equation}
where $C_0 (t)=x_0 (t) = C(t)$ is the correlation function itself. Each function satisfies
$C_n (0) = 1$ and $\Delta_{n+1}^2 C_{n+1}(t)$ acts as a memory kernel for the
dynamics of $C_n (t)$. The value of constants
$\Delta_{n+1}^2$ define  the value of the  memory kernels at time $t=0$. 

Both dynamical descriptions are equivalent \cite{moro1981} and as we show momentarily the constants $\Delta_n$ are equal to the Lanczos coefficients $b_n$.

Within the Mori formulation (\ref{mori}) the condition $b_{m+1} = 0$ is equivalent to the condition that the
normalized memory kernel $C_{m} (t)$ is a right-continuous 
Heaviside step function $\Theta(t)$, i.e., $C_{m}(t)$ decays arbitrarily slowly or
rather not at all. In this case, Laplace transforming (\ref{mori}) (assuming $m=3$ and $C_3 (t) =
\Theta (t)$) yields
\begin{equation}
\tilde{C}_0 (s) =
\frac{1}{s+\frac{\Delta_1^2}{s+\frac{\Delta_2^2}{s+\frac{\Delta_3^2}{s}}}} \, ,
\label{corrfrac}
\end{equation}
where $\tilde{C}_0 (s) = \tilde{x}_0 (s)$ is again the Laplace transform of the
autocorrelation function. 
Continued fraction representations in the context of Mori approach has 
been formulated in \cite{mori1965-1}. 

Comparing the Laplace transforms \eqref{LLT} and \eqref{corrfrac} readily
yields that both dynamical descriptions are equivalent and that $\Delta_n =
b_n$.

The recursive and Mori methods can both be understood from the lens of completely integrable dynamics, see \cite{Dymarsky_2020}.

As a spin-off observation, not related to the following
analysis, we note that the time evolution of $C_n (t)$ ($n
\geq 1$) corresponds to the time evolution of $x_n (t)$ but with a time
evolution operator obtained by deleting the first $n$ rows and columns of $L$.
(Also, the initial conditions are different,  $C_n (0)=1$ and $x_n(0)=0$ for $n\geq 1$.)

\section{Approximation of dynamics}
\label{Approx}

\subsection{Memory-kernel approximation}

In what follows we assume a thermodynamic limit such that $m$ is either exponentially large or infinite, rendering \eqref{LLT} and \eqref{corrfrac} infinite continuous fraction expansion.

Our main proposal is that exact  dynamics for $C_0 (t)$  may be reasonably 
approximated by a suitably ``truncated'' continuous fraction, not in
the sense that the resulting approximate correlation function is very close to $C_0(t)$ for all $t$, but that it has approximately the same early $t$ behavior and hence  equilibration time (defined below).

Our ``truncation'' scheme does not assume rendering the sequence of $b_n$ finite, but rather we propose to leave the first $R$ coefficients $b_n$ intact, while setting all subsequent Lanczos coefficients to be constant $b_n = b_R$ for all $n\geq R$. Although this may seem to be
a radical step from the Lanczos algorithm point of view, it yields a good approximation, as we see below.

From the Mori viewpoint, rendering all $\Delta_n=b_R$ constant for $n\geq R$ results in a successive application of
the same map 
\begin{equation}
\tilde{C}_{n-1}(\omega) = \frac{1}{i\omega + b_{R}^{2} \tilde{C}_{n}(\omega)}
\end{equation}
Here, instead of $s$ we used $s=a+i\omega$ and note that knowing $\tilde{C}_n(s)$ along the imaginary axis is sufficient to perform the inverse Fourier transform and determine $C_n (t)$. 
Successive application of the same transform makes all $\tilde{C}_n (\omega)$ for $n\geq R-1$ to be the same, and equal to a fixed point of the map 
$\tilde{C}_n (\omega)=\tilde{S}_{R}(\omega)$ for $n\geq R-1$, where 
\begin{equation}
\tilde{S}_{R}(\omega) = -i \frac{\omega}{2 b_{R}^2} +
\frac{1}{b_{R}}\sqrt{1-\left( \frac{\omega}{2 b_{R}} \right)^2}
\end{equation}
for $|\omega| \leq 2 b_{R}$, and
\begin{equation}
 \tilde{S}_{R}(\omega) = i\left(- \frac{\omega}{2 b_{R}^2} +
\text{sgn}(\omega)\frac{1}{b_{R}}\sqrt{\left( \frac{\omega}{2 b_{R}}
\right)^2 -1} \right)
\end{equation}
otherwise. Here $\text{sgn}(\omega)$ is the sign function.
This result already first appears in \cite{haydock1980}.

As a result we get the following approximation $\tilde{C}_R(\omega)$ to the original exact $\tilde{C}_0 (\omega)$: with all kernels $\tilde{C}_n (\omega)$ being equal to each other for $n\geq R-1$ the continued fraction expansion becomes finite, e.g.,
for $R=3$,
\begin{equation}
\tilde{C}_0^{R=3} (\omega) =
\frac{1}{i\omega+\frac{b_1^2}{i\omega+\frac{b_2^2}{i\omega+b_3^2 \tilde{S}_3
(\omega)}}}.
\label{capprox}
\end{equation}
Overall, this approximation is expected to be more accurate as $R$ increases. 
However, our main point is that in many practical situations it yields a reasonably accurate approximation already for small  $R$ ($\leq
10$). In such a case, the Fourier
transform of $\tilde{C}_0^R (\omega)$ can be readily evaluated 
numerically, giving rise to $C^R(t)$. The resulting
approximation is therefore determined by a very small number of Lanczos
coefficients, $b_n$ with $n \leq R$, but, as we see below, gives a good approximation to $C_0(t)$ for a wide range of $t$. 

\subsection{Equilibration time}
We define the equilibration time $\tau$ as the time when the
absolute value of $C(t)$ drops below some threshold value $g$ and never exceeds it again, i.e., $\left| C (t) \right| \leq g$ for $t\geq \tau$. Similarly, we denote the equilibration time of $C^R$ by $\tau_a$ (approximate equilibration time), 
$\left| C^{R} (t) \right| \leq g$ for $t\geq \tau_a$. Our definition
differs from other definitions of equilibration time in the literate, e.g.,~\cite{garcia2017}.
It is best suited for situations when $C(t)$ decays sufficiently fast. Physically, $\tau$ is characterizing local equilibration. 
Indeed in most cases $\tau$ is finite in the thermodynamic limit, while global thermalization time, i.e., diffusion time associated with $C(t)\sim t^{-1/2}$ in 1D systems would increase indefinitely with the system size. 

Put differently, the definition above  is not 
sensitive to functional behavior of $C(t)$ for large $t$, see the discussion in \cite{parker2019,viswanath2008}. Thus even if there is a slow decaying long tail,
 e.g.,~$\propto 1/t^\alpha$, $\alpha>0$, which is known to be sensitive to $b_n$ with very large $n$, the equilibration time $\tau$ may still be sensitive only to a handful of first $b_n$. 

Our approach is to estimate $\tau$ by evaluating $\tau_a$. To characterize the resulting discrepancy we introduce a relative error defined as 
\begin{equation}
\varepsilon_{\tau,R} = \left| \frac{\tau_a - \tau}{\text{min}(\tau_a,\tau)}
\right| \, .
\label{epstau}
\end{equation}

\subsection{Area estimate and equilibration time}
Since the evaluation of $\varepsilon_{\tau,R}$ is not possible without knowledge of the exact $\tau$, below we propose an indirect method to control the precision. Namely, we introduce another measure of precisions and to determine the minimal suitable value of $R$.

To assess the precision, we introduce the ``area measure'' defined as the area under $C^R(t)$,
\begin{equation}
\label{A}
 A_R = \int_0^{\infty} C^R (t) dt = \tilde{C}^R_0(0) \ .
\end{equation}
This measure is easily to calculate analytically, yielding 
\begin{equation}
A_R = \frac{1}{b_1} \Pi_{n=1}^{R-1} \left( \frac{b_n}{b_{n+1}} \right)^{\left(
-1 \right)^n} \, .
\label{area}
\end{equation}
In a similar way we introduce the area under the exact $C(t)$, $A= \tilde{C}_0(0)$.

The approximated correlation function $C^R(t)$ can only be a good description of  $C(t)$ 
so far the area measures for both are not significantly different from
each other. When $R$ increases, $A_R$ will converge to $A$ (this statement is intuitive but mathematically subtle, see \cite{viswanath1990}). 
We intend to choose $R$ as small as possible, so that $A_R$ will not be too different from $A$. As a measure of convergence we can use the relative error
\begin{equation}
\varepsilon_{A,R} = \frac{1}{2} \left( \frac{\left| A_R - A_{R+1} \right|}{A_R}
+ \frac{\left| A_R - A_{R+2} \right|}{A_R} \right) \, ,
\label{epsa}
\end{equation}
and choose the smallest $R$, for which $\varepsilon_{A,R}$
falls below some predefined threshold $a$.

We emphasize, that evaluating the area measure \eqref{area} is easy and only requires a small number of Lanczos coefficients.

As a consistency check of our proposal we note that for typical quantum systems satisfying the universal operator growth hypothesis of \cite{parker2019} and exhibiting linear growth of Lanczos coefficients $b_n=c\, n +d$, the area measure $A_R$ converges with the rate $1/R$. We consider an example of a system exhibiting such a behavior below. 

There are certainly examples where the area $A$ and hence the approximation (\ref{area}) 
diverges, e.g., autocorrelation functions of on-site spin operators in a diffusive $1$-d spin systems,
which decay as $\propto 1/\sqrt{t}$. In this case our area-based indirect measure of accuracy is not applicable, although the approach of estimating $\tau$ using $C^R (t)$ may still hold.

In the following, our goal is to demonstrate that there is a correlation between the relative
area error $\varepsilon_{A,R}$ and the relative equilibration-time error
$\varepsilon_{\tau,R}$ such that if $\varepsilon_{A,R}$ is reasonably
small, then $\varepsilon_{\tau,R}$ is relatively small as well, which in turn means that $\tau_a$ is a reasonable approximation for $\tau$. We show this numerically for several models with different archetypal dynamics.

\subsection{Example: SYK model}
First, to illustrate the approach above we consider a standard example of a quantum chaotic system, the SYK model describing $N$ interacting Majorana fermions \cite{Maldacena:2016hyu}.
This case can be treated semi-analytically. 
 In the large $q$ limit, where $q$ is the number of fermions in the interacting term, the correlation function and the Lanczos coefficients can be described analytically
\begin{equation}
\label{SYK}
C(t)={1\over \cosh^\eta (\alpha t)},\quad b_n^2=\alpha^2 n(n+\eta-1),
\end{equation}
with some positive $\alpha,\eta$. The same  $C(t)$ and $b_n$ appear in the context of 2d conformal field theories, with $\alpha$ being related to inverse temperature and $\eta$ to operator's dimension \cite{Dymarsky:2021bjq}. The correlation function \eqref{SYK} is decaying exponentially, 
with the ``area measure''  \eqref{A} 
\begin{equation}
A={\Gamma(1/2)\Gamma(\eta/2)\over 2\alpha\Gamma((1+\eta)/2)}
\end{equation} and thermalization time $\tau\approx -\ln g/(\alpha \eta)$. 

\begin{figure}[t]
\centering
\hspace{-1.0cm}
\includegraphics[width=0.75\columnwidth]{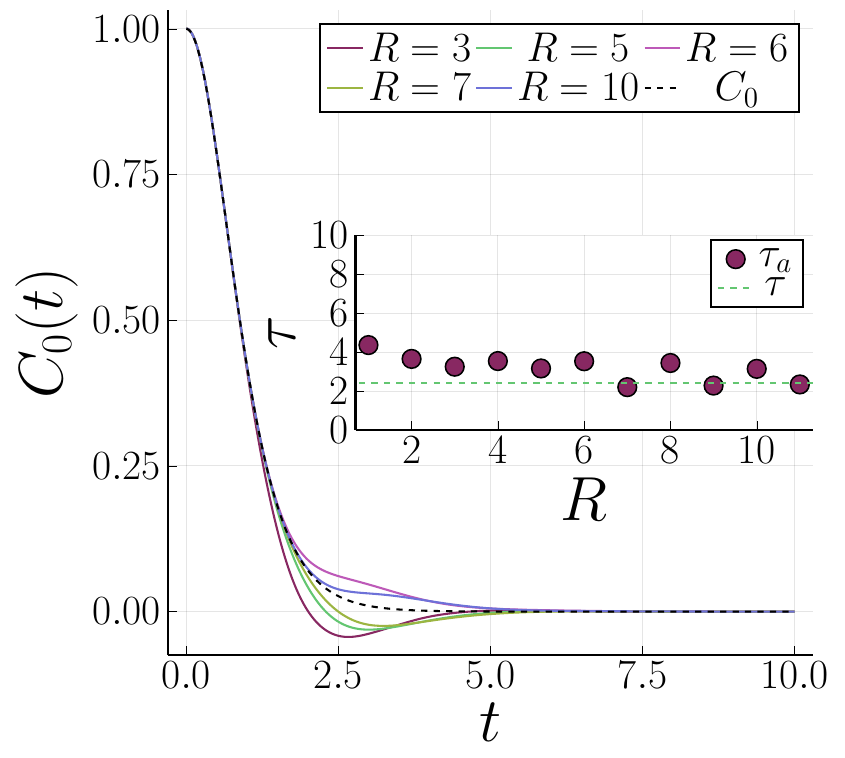}
\caption{Actual and approximated correlation functions for different values of $R$
for the SYK model with $\eta=2$. Inset shows 
equilibration times of approximated correlation functions for different values of $R$ with
equilibration-time threshold $g=0.03$.}
\label{SYKfigure}
\end{figure}

Now we use our approximation scheme by changing all $b_n$ for $n\geq R$ to be $b_n=b_R$. This behavior is very accurately describing actual Lanczos coefficients in a model of a 1D free scalar field, with a UV cutoff of order $b_R$, see \cite{Avdoshkin:2022xuw}. As one may expect, a UV cutoff does not significantly affect the 2-point function until the times inversely proportional to its value. Accordingly, the area measure of the new correlation function \eqref{area}, which can be evaluated analytically, converges to the true result with the $1/R$ rate, 
\begin{eqnarray}
\nonumber
A_R=\frac{\sqrt{\pi }\, \Gamma \left(\frac{\eta }{2}\right) \sqrt{\frac{\eta +R-1}{R}} \Gamma \left(\frac{R+1}{2}\right) \Gamma \left(\frac{R+\eta }{2}\right)}{2 \Gamma \left(\frac{\eta +1}{2}\right) \Gamma \left(\frac{R}{2}\right) \Gamma \left(\frac{1}{2} (R+\eta +1)\right)}=\\
A\left(1-\frac{1}{2 R}+\frac{2 \eta -1}{8 R^2}+\dots\right).
\end{eqnarray}
This confirms that taking $R\approx 10$ would provide a reasonably good accuracy for the estimation of the area measure. Similarly, $\tau_a$ will give a reasonable approximation to $\tau$, as follows from the numerical plot of $C^R (t)$ in Fig.~\ref{SYKfigure}.

\section{Numerics}
\label{Numerics}

\subsection{Tilted-field Ising model}

In the following, we test our approach numerically for several typical quantum models.
We begin with the tilted-field Ising model, with the Hamiltonian
\begin{eqnarray}
&& H = H_0 + B_x \sum_{l} \sigma_{x}^{l} \ , \\
&& H_0 = \sum_{l} J \sigma_{x}^{l} \sigma_{x}^{l+1} + B_z \sigma_{z}^{l} \ ,
\end{eqnarray}
where $\sigma_{x,z}$ are the respective spin components, $J$ is the spin
coupling constant, and $B_{x,z}$ are the components of the applied magnetic
field. We set $J=1.0$, $B_z=-1.05$ and use $B_x$ as tunable parameter. 
For $B_x = 0$ the magnetic field is
not tilted and the model is integrable, whereas for $B_x \neq 0$ the model
becomes non-integrable. As observable of interest, we consider a fast mode
\cite{Wang2022}
\begin{equation}
O \propto \sum_l \text{cos}(\pi l) h_l \, ,
\end{equation}
with $h_l$ being the local energy, i.e., $H=\sum_l h_l$, 
\begin{equation}
 h_{l}=J\sigma_{l}^{x}\sigma_{l+1}^{x}+\frac{B_{x}}{2}(\sigma_{l}^{x}+\sigma_{l+1}^{x})+\frac{B_{z}}{2}(\sigma_{l}^{z}+\sigma_{l+1}^{z}) \ .
\end{equation}
For the calculation, we
choose a system with $L=24$ spins and periodic boundary conditions. The dynamics of this observable is the
case of a rather fast decaying correlation function, for a further discussion see \cite{heveling2022}.
The finite length $L$ is only relevant for the numerical calculation of the 
actual correlation function.

\begin{figure}[t]
\centering
\hspace{-1.0cm}
\includegraphics[width=0.75\columnwidth]{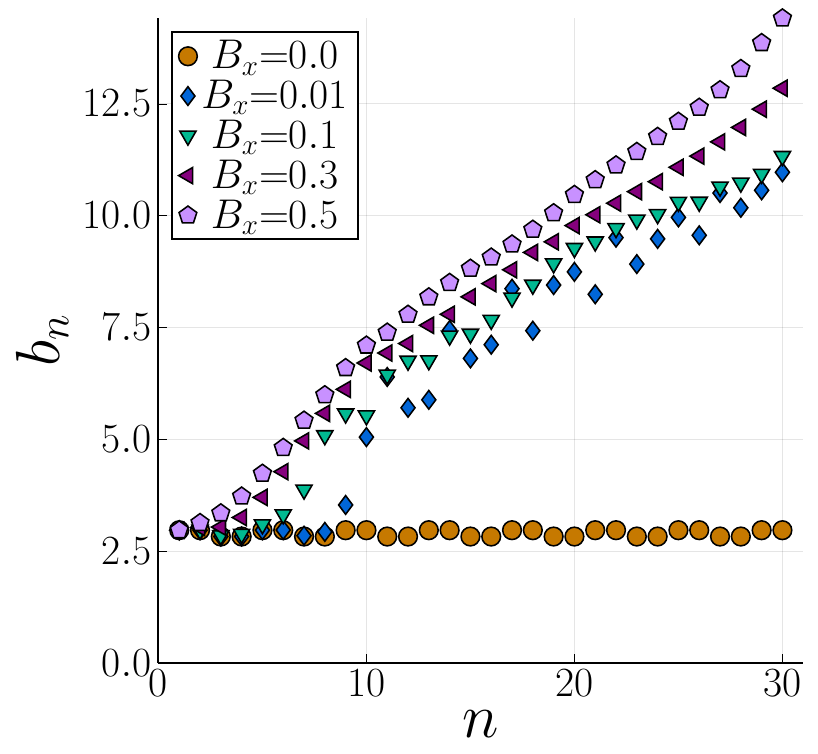}
\caption{Numerically calculated Lanczos coefficients for the tilted-field Ising
model.}
\label{lancztf}
\end{figure}

Corresponding Lanczos coefficients are depicted in Fig.\ \ref{lancztf}.
These coefficients are system size-independent. Their values are obtained by increasing the length
$L$ to be large enough such that finite size effects vanish. (For this model the first 
$L/2$ Lanczos coefficients correspond to infinite-system values.) 

\subsection{Spin-$1/2$ coupled to an Ising spin bath}

Next, we consider a model of a single spin-$1/2$ coupled to a bath, which
consists of an Ising spin model with an applied magnetic field. The
corresponding Hamiltonian reads
\begin{equation}
 H = H_S + \lambda H_I + H_B
\end{equation}
with
\begin{eqnarray}
&& H_S = \omega \sigma_z^S \  , \\
&& H_I = \sigma_x^S \sigma_x^0 \ , \\
&& H_B = \sum_{l} J \sigma_z^l \sigma_z^{l+1} +
B_x  \sigma_x^l + B_z  \sigma_z^l \ .
\end{eqnarray}
The parameters are chosen to be
\begin{equation}
 \omega = 1, \, J = 1, \, B_x = 1, \, B_z = 0.5 \, .
\end{equation}
The length of the bath is $L=20$ and we use periodic boundary conditions.
Finite $L$ effects are only important for the exact $C(t)$, all evaluated $b_n$ are universal. 
Here $H_S$ is the Hamiltonian of the
single spin, $H_B$ is the bath Hamiltonian, and $H_I$ introduces a
coupling of the single spin to the first spin of the bath via the
corresponding $x$ components. We investigate different scenarios for this model
by varying the coupling strength $\lambda$. As observables of interest for the
correlation function, we analyze here the $z$-component of the single spin,
$O=\sigma_z^S$, and the respective $x$-component, $O=\sigma_x^S$.

\begin{figure}[t]
\centering
\hspace{-1.0cm}
\includegraphics[width=0.75\columnwidth]{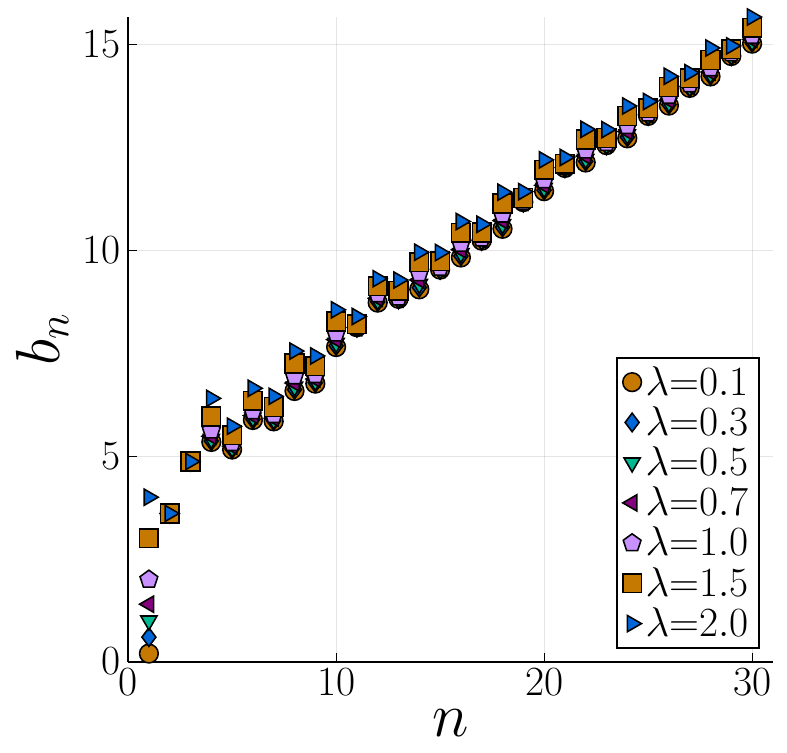}
\caption{Numerically calculated Lanczos coefficients for $\sigma_z^S$ in the model of a spin-1/2 coupled to a bath.}
\label{lanczsz}
\end{figure}

\begin{figure}[b]
\centering
\hspace{-1.0cm}
\includegraphics[width=0.75\columnwidth]{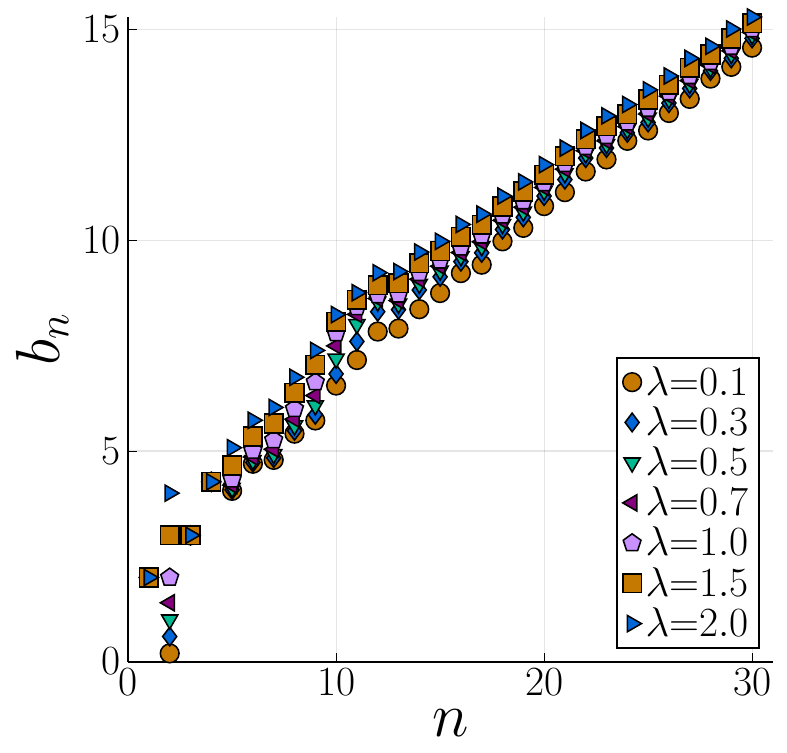}
\caption{Numerically calculated Lanczos coefficients for $\sigma_x^S$ the model of a spin-1/2 coupled to a bath.}
\label{lanczsx}
\end{figure}

The Lanczos coefficients are shown in Figs.\ \ref{lanczsz} and \ref{lanczsx}.
Note that for both observables all Lanczos coefficients, except for the first one,
are very similar. For the observable $\sigma_z^S$, the first Lanczos
coefficient $b_1$ is much smaller compared to the others $b_n$'s. It decreases for smaller interaction strength $\lambda$. For the
observable $\sigma_x^S$, similar behavior is exhibited by the second Lanczos coefficient $b_2$.
This feature has an immediate consequence for the value of $A_R$. 
For small coupling strengths $\lambda$, the
observable $\sigma_z^S$ exhibits a slow exponential decay
and $\sigma_x^S$ exhibits a slow, exponentially
damped oscillation \cite{DQT}, see Fig.\ \ref{dynbest}. 
Numerically calculated exact correlation functions for two sets of parameters are 
shown further below in Figs.\ \ref{dynbest},\ref{dynworst}. 
The curves are obtained using a numerical approach based on quantum typicality \cite{DQT}.

\subsection{Numerical equilibration times}

\begin{figure}[t]
\centering
\hspace{-1.0cm}
\includegraphics[width=0.75\columnwidth]{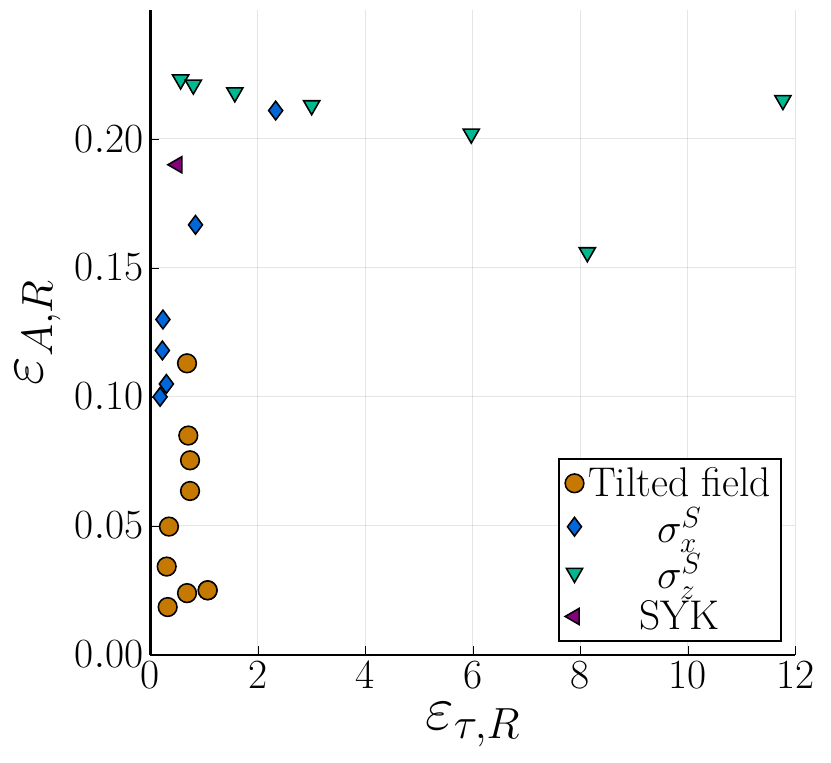}
\caption{Relative area error $\varepsilon_{A,R}$ (with threshold $a=0.25$)
versus relaxation time error $\varepsilon_{\tau,R}$ (with threshold $g=0.03$)
for all models considered. All $R$ lie between $1$ and $4$.}
\label{error1}
\end{figure}

\begin{figure}[t]
\centering
\hspace{-1.0cm}
\includegraphics[width=0.75\columnwidth]{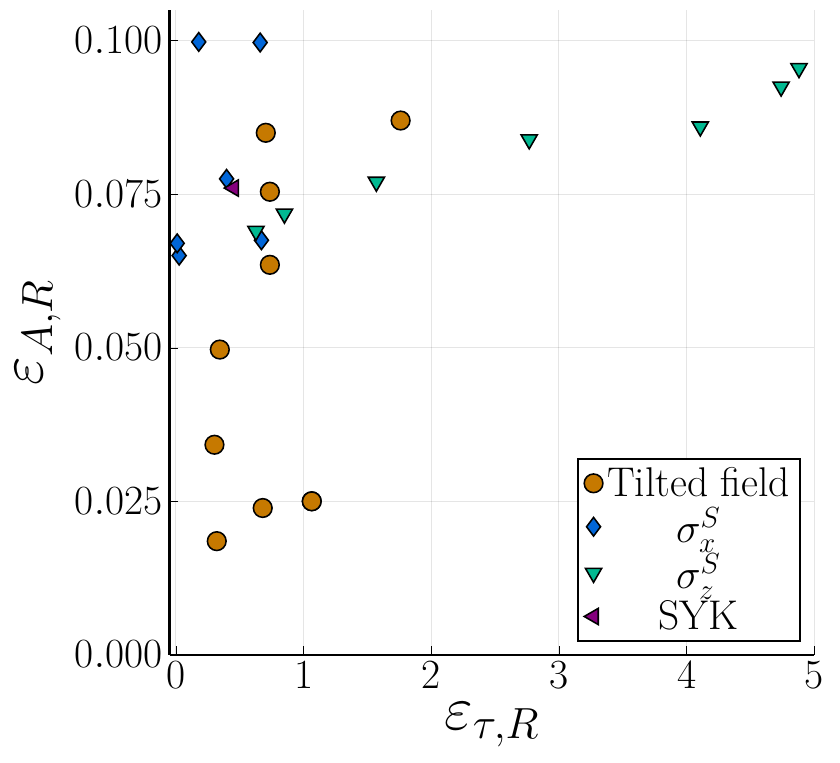}
\caption{Relative area error $\varepsilon_{A,R}$ (with threshold $a=0.1$)
versus relaxation time error $\varepsilon_{\tau,R}$ (with threshold $g=0.03$)
for all models considered. All $R$ lie between $1$ and $8$.}
\label{error2}
\end{figure}

We now study how the area estimation criterion can be used to determine the relevant $R$ to achieve a reasonable approximation of $\tau$. In figures \ref{error1} and
(\ref{error2}) we show the relaxation-time error $\varepsilon_{\tau,R}$ versus the
area error $\varepsilon_{A,R}$, for the area-error thresholds $a=0.25$ and $a=0.1$.
In both cases, we choose $g=0.03$ as threshold value to define equilibration 
 times $\tau$, $\tau_a$ from the numerically calculated curves
\cite{DQT}. While there is no functional dependence visible between
$\varepsilon_{A,R}$ and $\varepsilon_{\tau,R}$, one may conclude that, for all
cases shown, whenever the area error is below a reasonably small threshold, this assures that $\varepsilon_{\tau,R}$ is also sufficiently small, allowing to estimate $\tau$ from $\tau_a$, at lest within an order of magnitude. (A factor $12$ for $a=0.25$ and factor $5$ for
$a=0.1$.) Note that points corresponding to, e.g., the largest $R$ do not
systematically correspond to smallest $\varepsilon_{\tau,R}$ or
$\varepsilon_{A,R}$. Therefore, the area estimation appears to be a useful and
easily accessible tool to determine a minimal $R$ for a satisfactory
approximation of the correlation function in terms of a finite fraction 
(\ref{corrfrac}).


\begin{figure}[t]
\centering
\hspace{-1.0cm}
\includegraphics[width=0.75\columnwidth]{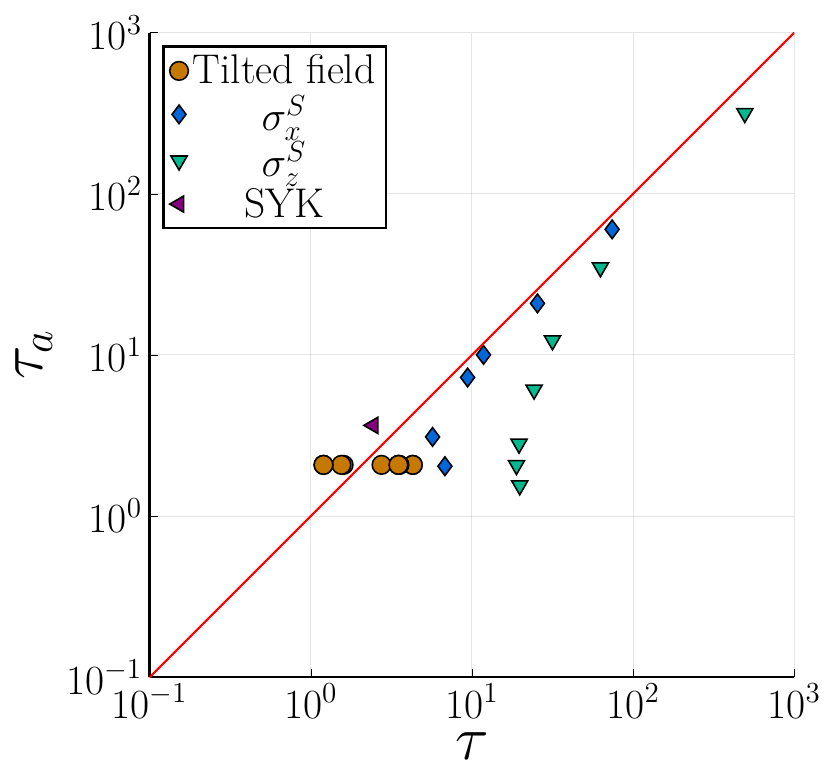}
\caption{Correct equilibration time $\tau$ versus approximated equilibration
time $\tau_a$, with $R$ from area threshold $a=0.25$ and equilibration-time
threshold $g=0.03$, for all models considered. All $R$ lie between $1$ and $4$.}
\label{times1}
\end{figure}

\begin{figure}[b]
\centering
\hspace{-1.0cm}
\includegraphics[width=0.75\columnwidth]{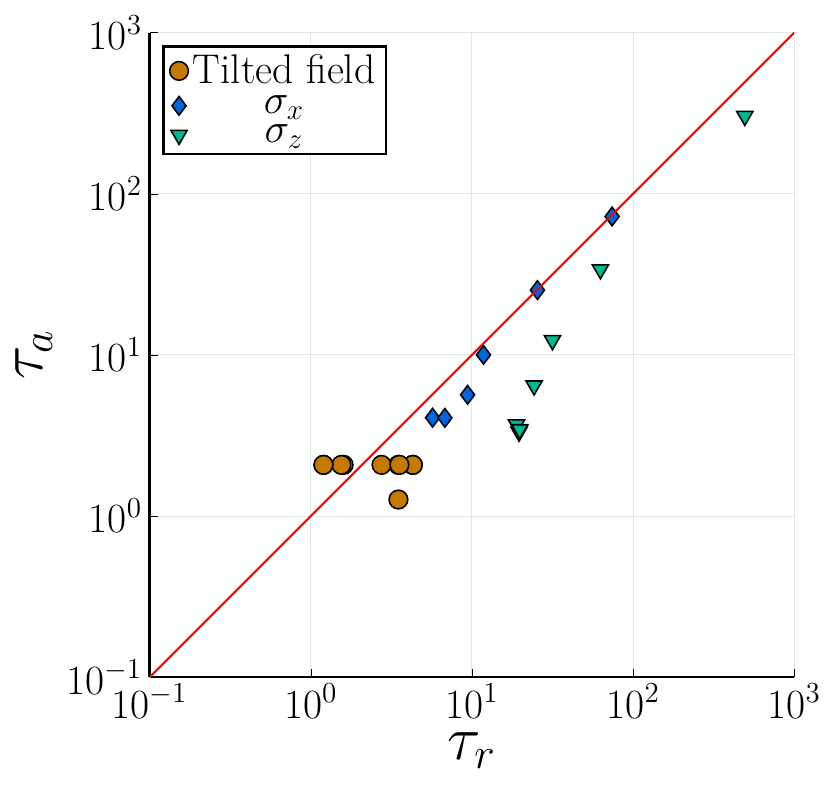}
\caption{Correct equilibration time $\tau$ versus approximated equilibration
time $\tau_a$, with $R$ from area threshold $a=0.1$ and equilibration-time
threshold $g=0.03$, for all models considered. All $R$ lie between $1$ and $8$.}
\label{times2}
\end{figure}

\begin{figure}[t]
\centering
\hspace{-1.0cm}
\includegraphics[width=0.75\columnwidth]{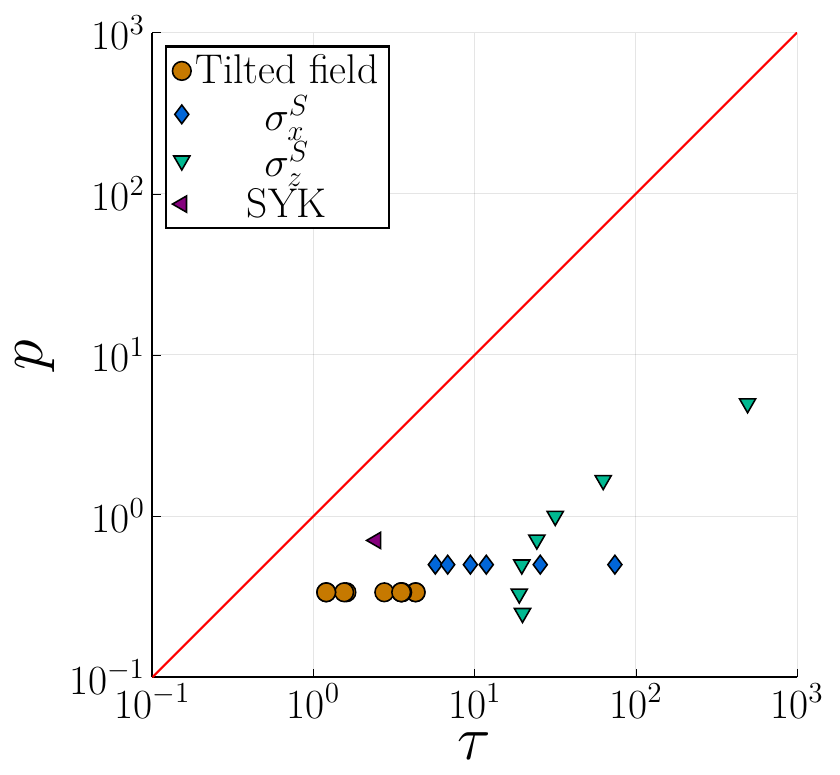}
\caption{Correct equilibration time $\tau$ versus bound $p$ from
\cite{garcia2017}, with equilibration-time threshold $g=0.03$, for all models
considered.}
\label{timesgarcia}
\end{figure}

Figures \ref{times1} and \ref{times2} show exact versus approximated
relaxation times for all models considered, again for the $R$ determined by the 
area threshold $a=0.25$ and $a=0.1$. As already visible in Figs.\
\ref{error1} and \ref{error2}, one finds that approximated and exact
relaxation times do not differ substantially from each other. Figures
\ref{times1} and \ref{times2} also show that our models cover a quite large
range of equilibration time scales, from fast decay for the tilted field model
to slow (exponential) relaxation and oscillation for the spin coupled to a bath
model \cite{DQT}. To further demonstrate the strength of our approach, we compare
our approximation of the equilibration time with an estimate of an upper bound
of the equilibration time derived in \cite{garcia2017}, which is depicted in
Fig.\ \ref{timesgarcia}. In \cite{garcia2017}, this bound $p$ is specified to
be proportional to $1/\sqrt{\partial^2 C_0 (t=0)/\partial t^2}$, which in the
Lanczos or Mori formulation amounts to $\propto 1/b_1$. (For simplicity, we
assume the proportionality constant to be $1$.) We also note that a somewhat similar approach to describe the autocorrelation function with help of the first few coefficients $b_n$  with the emphasis on $b_1$ was recently undertaken in \cite{zhang2023universal}. 

Although there are limitations to this comparison, partly because in
\cite{garcia2017} a rather different definition of equilibration time is
employed, one may observe that our estimated equilibration times are systematically much closer to the solid line, which represents the case when actual
equilibration time and the approximate one coincide. For the spin coupled to a bath model, the
approximation from \cite{garcia2017} yields a constant for the
$\sigma_x^S$ component and a square root dependence in the double logarithmic
plot for the $\sigma_z^S$ component, for the small-coupling/large
equilibration-time cases. Note that the slow oscillation for the $\sigma_x^S$
case \cite{DQT} is basically excluded in \cite{garcia2017}, because the
frequency distribution is not unimodal.


\begin{figure}[t]
\centering
\hspace{-1.0cm}
\includegraphics[width=0.75\columnwidth]{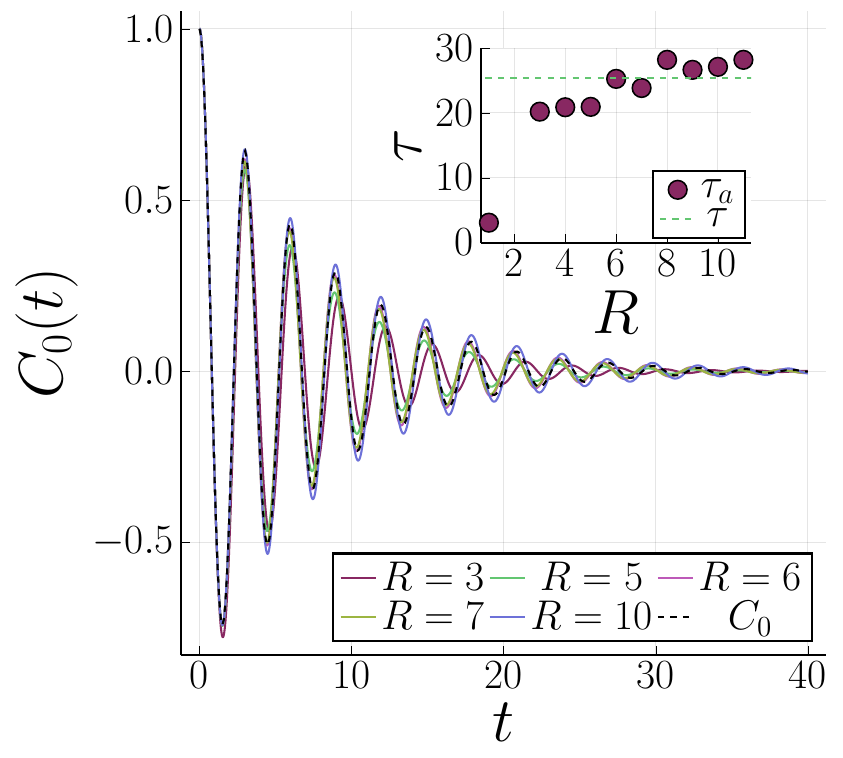}
\caption{Actual and approximated correlation functions for different values of $R$
for $\sigma_x^S$, $\lambda=0.5$ in the model of a spin-1/2 coupled to a bath. Inset shows 
equilibration times of approximated correlation functions for different values of $R$ with
equilibration-time threshold $g=0.03$.}
\label{dynbest}
\end{figure}

\begin{figure}[t]
\centering
\hspace{-1.0cm}
\includegraphics[width=0.75\columnwidth]{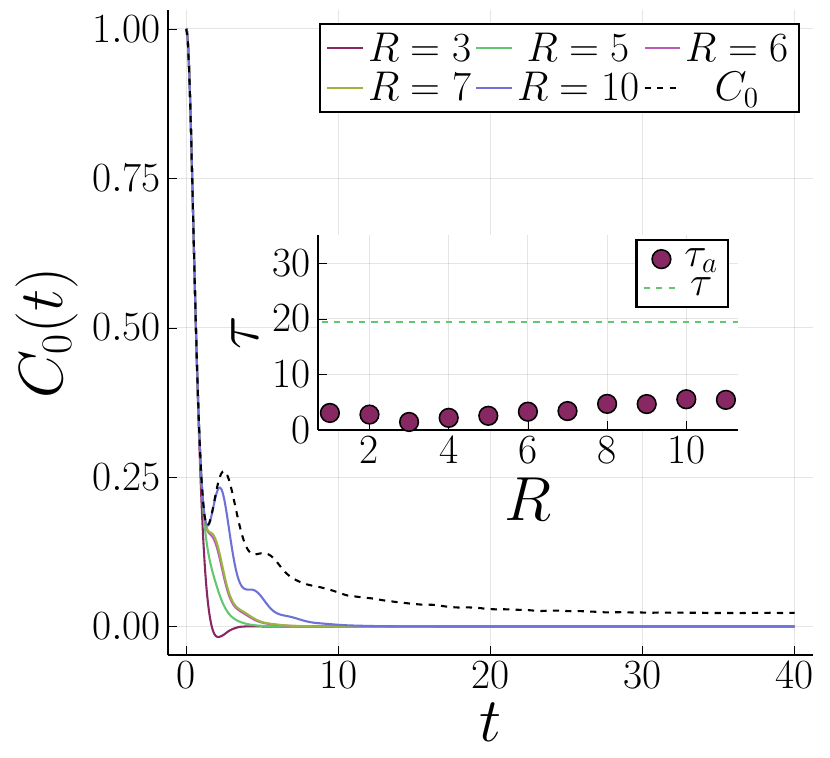}
\caption{Actual and approximated correlation functions for different values of $R$
for $\sigma_z^S$, $\lambda=1.0$ in the model of a spin-1/2 coupled to a bath. Inset shows 
equilibration times of approximated correlation functions for different values of $R$ with
equilibration-time threshold $g=0.03$.}
\label{dynworst}
\end{figure}

Beyond a mere estimation of equilibration time, we propose that the ``terminated''
continued fraction (\ref{capprox}) may give a reasonable approximation for the actual
correlation function itself. To analyze this feature, we show in Figs.\ \ref{dynbest},\ref{dynworst}
approximated correlation functions for different values of $R$ for examples that 
represent cases where our estimation for the equilibration time works best and worst,
i.e., $\sigma_x^S$ for $\lambda=0.5$ (best) and $\sigma_z^S$ for $\lambda=1.0$ (worst) 
for the model of a spin-1/2 coupled to a bath. 
(The examples are chosen according to the smallest
and largest relaxation time error $\varepsilon_{\tau,R}$ for area error threshold $a=0.1$ and
equilibration-time threshold $g=0.03$.)
One finds a very good agreement even for small values of $R$ for the good case, and 
there is no severe qualitative difference in the behavior in the worst case. For both cases one may 
deduce that the disagreement becomes smaller for growing $R$ and that there is a
convergence for large $R$, although this behavior is not robust for small $R$.

\section{Conclusion}

We have demonstrated that descriptions of the autocorrelation dynamics in the recursive (Lanczos algorithm) and Mori formulations are equivalent and, with help of Laplace transforms, can be recast in terms of a continued fraction expansion. 
Using this technique we introduced an approximation $C^R(t)$ 
to the autocorrelation function $C(t)$ that is fully determined by a small number of
the first $R$ Lanczos coefficients and therefore easy to evaluate numerically in most situations. We have compared suitably
defined equilibration times for an actual exact correlation function and the approximation, and found that the latter gives a reasonable estimation for the former. In order to control the precision of our approximation, we 
introduced an indirect ``area measure'' which evaluates the area under $C(t)$. 
This measure is very simple to evaluate and should be considered beforehand to determine the minimal necessary number $R$ of Lanczos coefficient, to achieve the desired level of accuracy.  We have numerically validated this approach by comparing how the quality of equilibration time estimation is correlated with the precision in estimating the area measure. We numerically considered several archetypal quantum models that exhibit a variety of different relaxation
behavior and a broad range of equilibration time scales.

We envision our approximation scheme for $C(t)$ in terms of $C^R(t)$ to be useful for various applications beyond an equilibration time estimation.

\section*{Acknowledgments}

This work has been funded by the Deutsche Forschungsgemeinschaft (DFG), under
Grants No. 397107022 (GE 1657/3-2), No. 397067869 (STE 2243/3-2), and No. 397082825, within the
DFG Research Unit FOR 2692, under Grant No. 355031190. 
AD is supported by the NSF under grants PHY-2310426.
We thank Lars Knipschild and
Robin Heveling for fruitful discussions.

\newpage

\end{document}